\documentclass[a4paper]{article}
\pdfoutput=1
\usepackage{xcolor}
\usepackage{enumitem}
\usepackage{amsmath}
\DeclareMathOperator*{\argmax}{arg\,max}
\usepackage{INTERSPEECH2020}
\usepackage[all]{nowidow}

\newcommand{\hadamard}{\odot}

\title{Deep Learning Based Open Set Acoustic Scene Classification}
\name{Zuzanna Kwiatkowska, Beniamin Kalinowski, Michał Kośmider, Krzysztof Rykaczewski}
\address{
  Samsung R\&D Institute\\
  Warsaw, Poland}
\email{\{z.kwiatkowsk, b.kalinowski, m.kosmider, k.rykaczewsk\}@samsung.com}

\begin{document}

\maketitle
\begin{abstract}
In this work, we compare the performance of three selected techniques in open set acoustic scenes classification (ASC). We test thresholding of the softmax output of a deep network classifier, which is the most popular technique nowadays employed in ASC. Further we compare the results with the Openmax classifier which is derived from the computer vision field. As the third model, we use the Adapted Class-Conditioned Autoencoder (Adapted C2AE) which is our variation of another computer vision related technique called C2AE. Adapted C2AE encompasses a more fair comparison of the given experiments and simplifies the original inference procedure, making it more applicable in the real-life scenarios. We also analyse two training scenarios: without additional knowledge of unknown classes and another where a limited subset of examples from the unknown classes is available. We find that the C2AE based method outperforms the thresholding and Openmax, obtaining $85.5\%$ Area Under the Receiver Operating Characteristic curve (AUROC) and $66\%$ of open set accuracy on data used in Detection and Classification of Acoustic Scenes and Events Challenge 2019 Task 1C.
\end{abstract}
\noindent\textbf{Index Terms}: acoustic scenes, open set classification, deep learning, conditional autoencoders

\section{Introduction}

A significant amount of research is still needed to reliably recognise sound in real-life scenarios, where multiple sounds are present, often simultaneously, and are distorted by the environment. Deep learning methods have been successfully used in addressing many of such audio processing problems, because they are suitable for solving complex and nonlinear learning tasks. Those methods are important from the academic point of view, but also find applications in the real-world use cases, considering a rapidly growing market of audio-related AI features~\cite{AAHearables2019}. One of the problems considered in audio AI is the problem of acoustic scene classification (ASC). 

ASC is a task of recognizing the environment of the recording device using only sound~\cite{Mcadams1993}. The challenges of such tasks depend on label granularity. If we consider a setting with broad scene classes, such as \textit{indoor}-\textit{outdoor}, the problem would be to obtain coherent class' features representations due to the high in-class data variability. When a finer granularity is considered, we usually focus on a selected subset of classes, but the models should still be able to react appropriately when the recording device is placed in an unknown environment. This creates a necessity to introduce a mechanism able to detect unknown examples and a need for improving methods concerning open set ASC. By an \textit{unknown}, we understand an example that does not belong to any of the classes that are recognizable by the given closed-set classifier. 

In an \textit{open set} classification problem, we aim to classify within a known set of labels (so called \textit{closed set}), but also to detect when a given sample is an unknown~\cite{TowardsOSRecog}. Although the problem was, so far, more researched in the computer vision (CV) field, it recently became popular in audio-related tasks. One of the reasons behind it was introducing the open set ASC as one of the problems in Task 1C of Detection and Classification of Acoustic Scenes and Events Challenge (DCASE 2019)~\cite{Mesaros2018_DCASE}. However, the methods proposed there have been very limited, and most utilized very similar techniques (see Section \ref{section:related_work}). Therefore, the main goal of this paper is to analyse the behaviour of open set classification methods that were designed for CV problems when applied to ASC.  

\section{Related Work}\label{section:related_work}

Many methods described in this section utilize the output of a closed set classifier as a part of their open set inference framework. In both ASC and CV, convolutional neural networks gained popularity due to their performance~\cite{dl_asc_review, review2, Dhillon2019}. In our experiments, we also used them as the open set methods' core~\cite{SCModel}.

In terms of the open set classification in CV, the first paper formalizing the problem was the work of Scheirer et.\ al.~\cite{TowardsOSRecog}, where a simple SVM-based method was proposed and tested on the ImageNet~\cite{imagenet_cvpr09}. A direct extension of this approach called W-SVM~\cite{prob_mod} additionally consisted of the Extreme Value Theory (EVT) outlier fitting as a method for calibrating scores obtained from a closed set classifier. Since then, the EVT became widely adopted~\cite{jainMulticlassOpenSet2014, zhangSparseRepresentationBasedOpen2017}. However, those methods are specific to SVM and are not easily applicable to deep learning.

More recently Openmax~\cite{Openmax} was proposed, which also relies on EVT. It modifies softmax to accommodate the unknown class. It is also specifically designed for deep learning. The idea of this technique was to construct a feature representation of each class and measure how much new examples diverge from these representations. When the divergence was substantial, a probability that a new example was from a predicted class was penalized, in favour of the class representing the unknowns. An idea to extend Openmax with a feature representation produced by an autoencoder was introduced in the Classification-Reconstruction Learning~\cite{crosr}.

In Class-Conditioned Autoencoder (C2AE)~\cite{c2ae}, the autoencoder was trained to reconstruct examples correctly if they were conditioned with a correct label, and incorrectly otherwise. During the inference, the reconstruction errors from all known labels were used to detect if the considered example diverges from the training data. More methods concerning open set classification in CV can be found in~\cite{Geng_2020}. 


In ASC, methods used so far were mostly limited to thresholds~\cite{first_1c_2019, second_1c_2019, 4_1c_2019, 5_1c_2019}, meaning the value $\varepsilon \in (0,1)$ was chosen and the example was predicted as unknown if the predicted class probability was lower than $\varepsilon$. Although in the DCASE 2019 Challenge they performed relatively well, the main concern was the neural networks' robustness problem, where an incorrect classification may have a relatively high confidence. Similarly to the CV field, there were also attempts to use both SVM-~\cite{osasc_method1} and autoencoder-~\cite{third_1c_2019} based methods, but, so far, with worse results than thresholds.

Taking into account the similarity between the closed set deep learning methods used in CV and audio processing, the question arises if using the open set algorithms developed in CV would produce an improvement in ASC over methods used so far. 

\section{Selected Methods}

For experimental purposes, we decided to choose three open set classification techniques: the thresholds, Openmax and the Adapted C2AE (our variant of C2AE). The thresholds were chosen because, so far, they were used most often in ASC tasks and because it is the most straightforward and an easily reproducible technique. The Openmax algorithm was chosen as an example of the EVT fitting. We also wanted to evaluate a method based on autoencoders, due to their popularity in anomaly detection, and therefore decided to choose C2AE which is relatively new.

All of these techniques utilize the output of a closed set classifier. To allow for a more fair comparison of the selected methods, and not of the underlying classifiers, we first trained a common base model.

\subsection{Closed Set Classifier Base}
In our experiment, by the classifier we understand a deep neural network with a softmax output. We selected two approaches to training the classifier:
\begin{enumerate}[label=(C\arabic*)]
    \item \label{classifier_method_1} the classifier is trained only on the target classes,
    \item \label{classifier_method_2} the classifier is trained on the target classes and an additional class with some examples of unknowns.  
\end{enumerate}

In the \ref{classifier_method_1} approach, we need to separate the known subspace from the unknown one without any prior knowledge of the unknown subspace. In this case, the challenge is to classify correctly the unknown examples that are very similar to the target ones (for example if we had \textit{metro} vs. rest classification, we could have difficulties with rejecting \textit{tram} as an unknown class).

The \ref{classifier_method_2} approach can be beneficial, as we also model some part of the unknown examples subspace. We may choose unknown examples that are very similar to our target classes in terms of feature representations, which may allow more precise separation. However, in many real-life use cases, we do not have the data with unknown examples. Moreover, such unknown class may have a high in-class examples variability, which could lead to poorer classifier performance on this class. 

\subsection{Thresholds}
Using thresholds on the softmax output of a closed set classifier is the most straightforward approach to enabling the usage of such models in solving an open set problem. However, it is based on an assumption that the correctness of prediction is correlated with the softmax confidence output, which may not always be the case (an example of such experiment can be seen in~\cite{adv}).

Let us denote by $\hat{y}$ a softmax output of a closed set classifier for a given example and by $\varepsilon$ an assumed decision threshold. Then we classify an example as
\begin{equation}
    \begin{cases} 
        \text{unknown,} & \text{if } \max\limits_{1 \leq i \leq K}(\hat{y}_i) < \varepsilon, \\
        \argmax\limits_{1 \leq i \leq K}(\hat{y}_i), & \text{otherwise,} 
  \end{cases}
\end{equation}
for $\varepsilon \in (1/a, 1)$, where $K$ is the number of classes. Note that in \ref{classifier_method_2}, the unknown class can be also predicted in the second case, because it is one of the classes that the classifier is trained on. Two out of DCASE 2019 Task 1C winning solutions were utilizing \ref{classifier_method_2} with thresholds method.

\subsection{Openmax}


The main idea of Openmax is to use EVT to extend softmax with an additional value representing the unknowns. Its meaning can be interpreted using notions of an outlier and a divergence. A divergence is defined as a weighed sum of Euclidean distance and cosine similarity. Outliers of a given class are defined as points diverging from the average logits (values just before the softmax activation) for examples of this class. Given this, the additional value represents the probability that the predicted logits are outliers from all of the classes. 

To estimate the probability that the prediction is an outlier, $K$ Weibull distributions are fitted to represent the maximal possible divergences from the average logits for each of the known classes. The probability that the maximum is smaller than the divergence for the considered example is used as the probability of being an outlier. The distributions are fitted solely using correctly classified examples for each of the classes from the training set. The training and inference procedures are shown in Algorithm 1 and 2 in~\cite{Openmax}. 

\subsection{Adapted C2AE}

The C2AE model is based on the idea of class conditioning in the latent space of an autoencoder. The autoencoder is trained to reconstruct an example correctly when it is conditioned with its ground truth label, and incorrectly otherwise. Originally, the incorrect reconstruction means reconstructing the example into one from a different class, e.g. a \textit{metro} into \textit{office}. 

The original C2AE training procedure begins with the encoder and classifier being trained jointly for the task of a closed set classification on some $K$ classes. Then, the encoder (frozen) and decoder are trained for the task of reconstruction. When the training is finished, the reconstruction errors from correct and incorrect conditioning are collected for the training set examples. Based on both types of errors, the threshold is found to differentiate between good and bad reconstruction. The algorithm of finding such threshold is described in Section 3.2.3 of the original paper~\cite{c2ae}. During the inference, a prediction is obtained from a classifier and the $K$ reconstruction errors are obtained by conditioning a new example with each of the known classes. If the minimum of reconstruction errors is lower than a previously computed threshold, the prediction is adopted. Otherwise, it is rejected and the example is predicted as an unknown.

In order to utilize our base classifier in C2AE, we decided to introduce the Adapted C2AE. In terms of training, we separately trained a classifier and the entire autoencoder, so a classifier does not need to be crafted specifically for this task (i.e. include an encoder). Based on the results in the original paper, we also decided to choose the threshold manually instead of automatically, to simplify the implementation and because the gain from choosing it automatically was not substantial (see Figure 4 in~\cite{c2ae}). 

In terms of the inference, we decided to condition the autoencoder with the output of a classifier and not all labels. This change was not significant, if the example considered in the inference was either an unknown or indeed from the predicted class. However, if the example was not from the predicted class, but still from some other known class, the minimum of the reconstruction errors would be small and the original C2AE would misclassify the example as an incorrect known class. Our inference would raise the unknown class in such case. From the practical point of view, the classification would still be incorrect in both cases, and therefore we decided to change the conditioning as it resulted in a faster inference procedure.

We also redefined what the incorrect reconstruction means by reconstructing the example into silence (zeros), instead of an example from another class, when conditioned with a wrong label. By choosing a constant, we aimed to stabilize the training.

\section{Experiment Setup}

\subsection{Data and Processing}

In the experiments, we used the \textit{TAU Urban Acoustic Scenes 2019 Open set development dataset}~\cite{Mesaros2018_DCASE} (\textit{TAU Open Set}) that was also used in DCASE 2019 Task 1C. The organizers did not provide the labels for the leaderboard and the evaluation datasets after the contest ended. Therefore, for our experiments, we use the official training and testing subsets of the development dataset. We additionally separated 10\% of examples for tuning purposes, maintaining the distribution of classes. The dataset consisted of 10 known classes (used in \ref{classifier_method_1} and \ref{classifier_method_2} training) and 1 unknown class (used only in \ref{classifier_method_2}). We did not augment the dataset.

Each example was a mono recording with 48 kHz sampling rate and of 10 seconds' length. We processed each example into a logarithmically scaled mel-spectrogram with 256 bins using Short Time Fourier Transform with a window of size 2048 samples and 512 samples overlap. Moreover, for each bin, we calculate its mean and standard deviation across all examples, and we standardize each bin with those values. This results in the output of size $862 \times 256$, the former being a time dimension, and the latter a frequency dimension.

\subsection{Deep Learning Models}
In the experiment, the two neural networks were trained: a classifier as a base model, and an autoencoder for Adapted C2AE inference. For models training and inference implementations we used TensorFlow~\cite{tensorflow2015-whitepaper}. We implemented Openmax based on the code provided in~\cite{Openmax_repo}. 

\subsubsection{Classifier}
We decided to use a small model (70k parameters) with a straightforward training procedure to allow for an easier reproduction of the obtained results. The model has already proved successful in ASC~\cite{SCModel, SCModel2}. The classifier architecture is described in Table~\ref{tab:model}. The classifier outputs a vector of length $K=10$ for \ref{classifier_method_1} and $K=11$ for \ref{classifier_method_2}.

\begin{table}[th]
    \caption{The architecture of the classifier.}
    \label{tab:model}
    \centering
    \begin{tabular}{lrrr}
        \toprule
        \multicolumn{1}{c}{\textbf{Layer}} & \textbf{Outputs} & \textbf{Kernel} & \textbf{Stride} \\ 
        \midrule
        \textit{Conv2D+ReLU+BN} & 16 & 3 & 1 \\
        \textit{Conv2D+ReLU+BN} & 32 & 3 & 2 \\
        \textit{Conv2D+ReLU+BN} & 32 & 3 & 1 \\
        \textit{Conv2D+ReLU+BN} & 64 & 3 & 2 \\
        \textit{Conv2D+ReLU+BN} & 64 & 3 & 1 \\
        \textit{AveragePooling} & 64 & - & - \\
        \textit{Dense+Softmax} & $K$ & - & - \\ 
        \bottomrule
    \end{tabular}
\end{table}

The model utilizes the Batch Normalization technique~\cite{batchnorm}. Both \ref{classifier_method_1} and \ref{classifier_method_2} were trained using GPU for 100 epochs with a mini batch of size 32, using categorical cross entropy loss. As the final model we used the best one in terms of loss on the validation dataset. Data was shuffled before each epoch. We used the \textit{Adam} optimizer~\cite{adam} with $\alpha = 0.001$. 

\subsubsection{Autoencoder} 

In the autoencoder architecture, we decided to use convolutional layers, which proved useful in the tasks related to acoustic scenes, as well as dense layers, to guarantee the data compression in the latent space. As a conditioning layer, we used the FiLM layer~\cite{Perez2018FiLMVR}, as proposed in the original C2AE paper~\cite{c2ae}, which is the following linear transformation on the latent space   

\begin{equation}
    \label{eq:film}
    o = \mathcal{H}_\alpha(y) \hadamard z + \mathcal{H}_\beta(y) 
\end{equation}
where $\mathcal{H}_\alpha$ and $\mathcal{H}_\beta$ are dense layers with the output size matching that of the latent space, $z$ is the latent space representation of a given example and $y$ is the one-hot encoded label of that example (1 is used for positive and -1 for negative class). The $\hadamard$ operation is the Hadamard product. A full architecture is shown in Table \ref{tab:autoencoder}.  

\begin{table}[th]
    \caption{The architecture of the autoencoder.}
    \label{tab:autoencoder}
    \centering
    \begin{tabular}{ll}
        \toprule
        \multicolumn{1}{c}{\textbf{Layer}} & \textbf{Outputs} \\ 
        \midrule
        \textit{Conv2D+ReLU$^*$} & 16 \\
        \textit{Conv2D+ReLU$^*$} & 8 \\
        \textit{Conv2D+ReLU$^*$} & 4 \\
        \textit{Dense} & 512 \\
        \textit{Dense} & 128 \\
        \textit{FiLM Layer} & 128 \\
        \textit{Dense} & 128 \\
        \textit{Dense} & 512 \\
        \textit{Dense} & 1224 \\
        \textit{Conv2DTranspose+ReLU$^*$} & 4 \\
        \textit{Conv2DTranspose+ReLU$^*$} & 8 \\
        \textit{Conv2DTranspose+ReLU$^*$} & 16 \\
        \bottomrule
        $^*$ With kernel and stride 3.
    \end{tabular}
\end{table}

Similarly to the classifier, the autoencoder was trained using a single GPU for 100 epochs with a mini batch of size 32. For each example in one batch, both reconstructions (conditioned correctly and incorrectly) were performed. As a loss for those two tasks we used a weighed sum of Mean Square Errors with weight $0.8$ for the correct reconstruction and $0.2$ for the incorrect one. As the final model we used the best one in terms of the loss on the validation dataset. The dataset was shuffled before each epoch, and we used \textit{Adam} optimizer with $\alpha = 0.001$. The autoencoder had approximately $1.4M$ parameters. In the autoencoder fitting procedure, the final threshold was found manually and set to $0.3$. The reconstruction errors were calculated using Mean Absolute Error.

\section{Results}
\subsection{Validation Metrics}
As a first validation metric, we used the metric from DCASE 2019 Task 1C challenge which was a weighed sum of known and unknown accuracy, denoted $\mathrm{ACC}_{K}$ and $\mathrm{ACC}_{U}$, respectively. In order to calculate the score, first, the accuracy for each class, including unknown, was computed. Then, as $\mathrm{ACC}_{K}$, we took an average of per-class accuracies of the known classes and as $\mathrm{ACC}_{U}$ we took the accuracy for the unknown class. The final score was computed per the formula 
\begin{equation}
    \mathrm{ACC} = 0.5 \cdot \mathrm{ACC}_{K} + 0.5 \cdot \mathrm{ACC}_{U}.    
\end{equation}
For calculation we used the toolbox provided by the DCASE 2019 organizers~\cite{sed_eval}. However, we are aware of the disadvantages of this metric, such as an arbitrary choice of weights and susceptibility to fluctuations, and therefore we decided to use Area Under ROC (AUROC) metric as well.  

\subsection{Discussion}

\begin{table}[th]
  \caption{DCASE 2019 score on \textit{TAU Open Set} test set.}
  \label{tab:results_tut}
  \centering
  \begin{tabular}{lrrr}
    \toprule
    \multicolumn{1}{c}{\textbf{Model}} & \textbf{$\mathrm{ACC}_{K}$} & $\mathrm{ACC}_{U}$ & $\mathrm{ACC}$ \\
    \midrule
    \textit{\ref{classifier_method_1} closed set} & 71.5\% & - & -\\
    \midrule
    \textit{\ref{classifier_method_1} with $\varepsilon = 0.5$} & 63.8\% & 35.9\% & 49.9\%\\
    \textit{\ref{classifier_method_1} with $\varepsilon = 0.6$} & 55.3\% & 56.5\% & 55.9\%\\
    \textit{\ref{classifier_method_1} with $\varepsilon = 0.7$} & 45.8\% & 76.2\% & 61.0\%\\
    \textit{\ref{classifier_method_1} with Openmax} & 46.7\% & 60.9\% & 53.8\%\\
    \textit{\ref{classifier_method_1} with Adapted C2AE} & 60.2\% & 70.4\% & 65.3\%\\
    \midrule
    \textit{\ref{classifier_method_2} with $\varepsilon = 0.5$} & 65.9\% & 33.3\% & 49.6\%\\
    \textit{\ref{classifier_method_2} with $\varepsilon = 0.6$} & 57.3\% & 49.3\% & 53.3\%\\
    \textit{\ref{classifier_method_2} with $\varepsilon = 0.7$} & 48.8\% & 66.4\% & 57.6\%\\
    \textit{\ref{classifier_method_2} with Openmax} & 38.9\% & 74.8\% & 56.8\%\\
    \textit{\ref{classifier_method_2} with Adapted C2AE} & 59.2\% & 72.8\% & 66.0\%\\
    \bottomrule
  \end{tabular}
\end{table}

The $\mathrm{ACC}$ results are shown in Table \ref{tab:results_tut}. The most surprising is that the \ref{classifier_method_2} methods, that utilize examples of unknowns, do not provide improvements over their \ref{classifier_method_1} counterparts. All approaches experience reduced accuracy on the known classes. This is expected due to the possibility of mistaking examples for unknowns. Openmax, despite its complexity, has lower accuracy than thresholding for unknowns at the same accuracy for known classes. The best results were reached using the Adapted C2AE, with a relatively small deterioration in known accuracy in relation to the gain obtained in the unknown classification.

\begin{figure}[t]
  \centering
  \includegraphics[width=\linewidth]{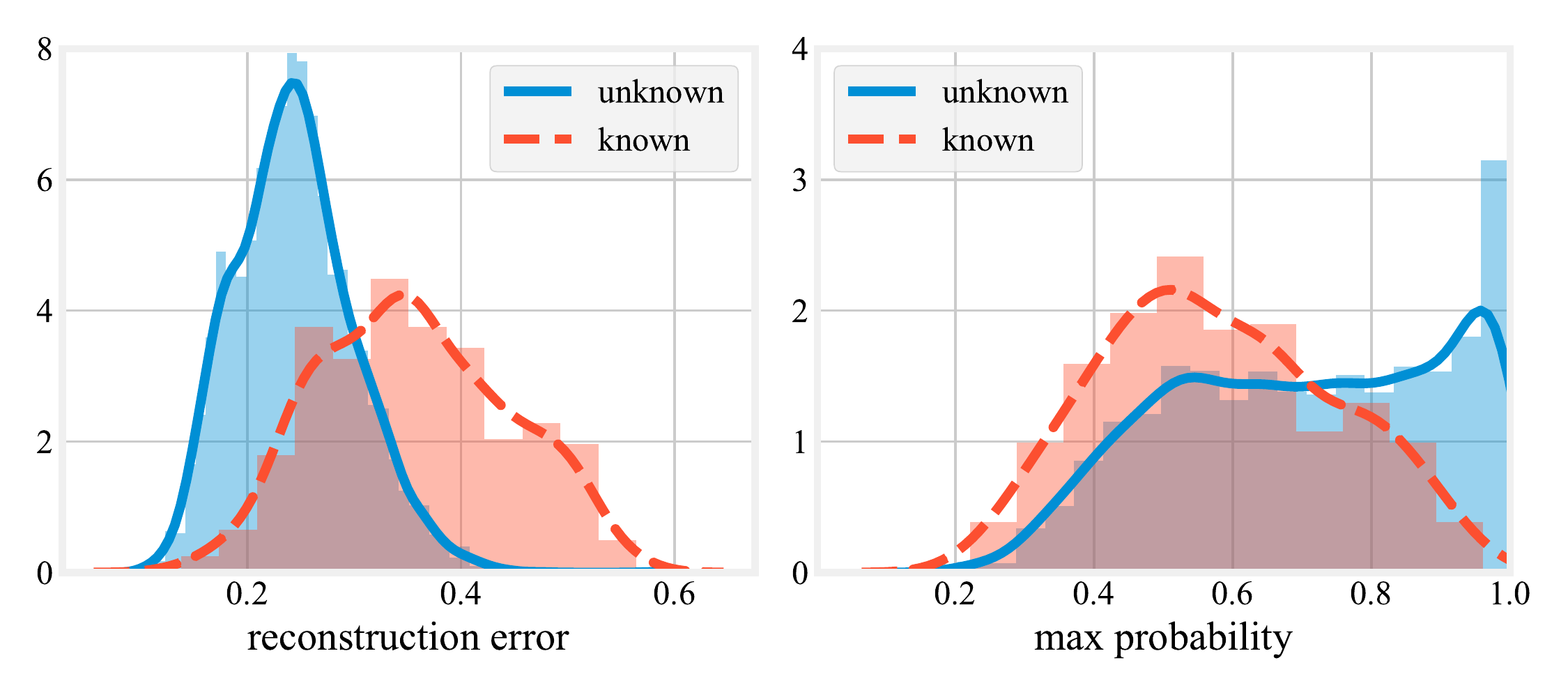}
  \caption{The comparison of known and unknown distributions separation for C2AE (left) and thresholding on softmax (right) on TAU Open Set test set.}
  \label{fig:rec_errors}
\end{figure}

Moreover, in Figure \ref{fig:rec_errors}, we present a comparison of known and unknown distributions' separation. Note that the $x$ axis is different in both subplots, because the threshold and separation is based on the predicted class probabilities in the softmax thresholding method, while in C2AE it is based on the reconstruction errors. It can be observed, that the intersection of knowns and unknowns in C2AE is smaller than the one based on probability, which may imply that the separation based on reconstruction error is more stable and produces better differentiation between known and unknown examples.  

Because there was no substantial gain in the accuracy between \ref{classifier_method_2} and \ref{classifier_method_1}, we calculated the AUROC scores only for \ref{classifier_method_1} training variant. AUROC was computed for unknown-vs-known classification by changing the thresholds. The results are shown in Table \ref{tab:tut_AUROC}.

\begin{table}[th]
  \caption{AUROC metric for \ref{classifier_method_1} on \textit{TAU Open Set} test set.}
  \label{tab:tut_AUROC}
  \centering
  \begin{tabular}{rccc}
    \toprule
    \textbf{Model} & Thresholds & Openmax & Adapted C2AE \\
    \midrule
    \textbf{AUROC} & 68.7\% & 71.1\% & 85.5\% \\
    \bottomrule
    \vspace{-0.95cm}
  \end{tabular}
\end{table}

From AUROC we can conclude that the Openmax works better than a simple thresholding technique. The initial advantage of thresholds over the Openmax in terms of $\mathrm{ACC}$ could be the result of $\mathrm{ACC}$ favouring better score on $\mathrm{ACC}_{U}$, which can be artificially increased by manipulating the threshold $\varepsilon$ value. From the qualitative point of view, implementing and understanding the Openmax and its EVT grounds requires more time and expert knowledge than a straightforward thresholding method, and due to the negligible performance difference between them, thresholds could be a simpler, yet still effective choice. However, overall, the performance of Adapted C2AE is still much better than of the previous two, which confirms the results obtained through $\mathrm{ACC}$.  

There are multiple ways how the experiments proposed in this paper could be extended. In terms of test data, it could be beneficial to run similar experiments on a completely different dataset that partially includes the same labels, such as LITIS Rouen \cite{litis}. This would require applying additional techniques for classification with mismatched recording devices, because different devices could lower the classifier performance and distort the obtained results. Moreover, the \ref{classifier_method_2} classifier could be extended to use multiple unknown classes instead of just one, aggregating examples with similar category, for example \textit{unknown transport}, \textit{unknown indoor}, etc. Both classifiers could also be extended to a multi-label instead of a single-label classification. Finally, in this paper we evaluated discriminative methods, because we wanted to compare them in a common training framework, but it could be beneficial to extend this framework to generative methods as well (e.g.\ the G-Open\-max~\cite{ge2017generative}). 

\section{Summary}
In this paper, we presented the results for open set ASC using three techniques. The first one was applying thresholds to the softmax output of a neural network classifier, which is the classic approach used so far in ASC. The second one, called Openmax, was a popular method derived from the CV field. The third one, we proposed the Adapted C2AE, which is a modified version of C2AE, where we applied changes allowing a more fair comparison with the other methods. Adapted C2AE outperformed both thresholding and Openmax, which shows that the autoencoder based solutions may be promising in the open set ASC.

\clearpage

\bibliographystyle{IEEEtran}
\bibliography{mybib}

\end{document}